\documentclass[pra,twocolumn,longbibliography]{revtex4}

\usepackage[utf8]{inputenc}
\usepackage{bm}
\usepackage{mathtools}
\usepackage{bbold}
\usepackage{amsmath}
\usepackage{amssymb}
\usepackage{hyperref}
\usepackage[normalem]{ulem}
\usepackage{color}
\usepackage[dvipsnames]{xcolor}

\begin{document}

\title{Quantum degeneracy and spin entanglement in ideal quantum gases}
\author{Fatma Zouari Ahmed}
\affiliation{Universit\'{e} El-Oued 39000, and LRPPS Universit\'{e} Kasdi Merbah,
Ouargla, 30000, Algeria}
\author{Mohammed Tayeb Meftah}
\affiliation{Universit\'{e} Kasdi Merbah, Ouargla, 30000, Algeria}
\author{Tommaso Roscilde \footnote{Corresponding author: tommaso.roscilde@ens-lyon.fr .}}
\affiliation{Univ Lyon, Ens de Lyon, CNRS, Laboratoire de Physique, F-69342 Lyon, France}


\begin{abstract}
Quantum degeneracy is the central many-body feature of ideal quantum gases stemming from quantum mechanics. In this work we address its relationship to the most fundamental form of non-classicality in many-body system, i.e. many-body entanglement. We aim at establishing a quantitative link between quantum degeneracy and entanglement in spinful ideal gases, using entanglement witness criteria based on the variance of the collective spin of the spin ensemble. We show that spin-1/2 ideal Bose gases do not possess entanglement which can be revealed from such entanglement criteria. On the contrary, ideal spin-1/2 Fermi gases exhibit spin entanglement revealed by the collective-spin variances upon entering quantum degeneracy, due to the formation of highly non-local spin singlets. We map out the regime of detectable spin entanglement for Fermi gases in free space as well as in a parabolic trap, and probe the robustness of spin entanglement to thermal effects and spin imbalance. Spin entanglement in degenerate Fermi gases is amenable to experimental observation using state-of-the-art spin detection techniques in ultracold atoms. 
\end{abstract}
\maketitle

\section{Introduction: spin entanglement of identical particles}
The indistinguishability of identical quantum particles has fundamental implications for the many-body physics of ideal quantum gases. Even in the absence of interactions, quantum statistics alone (bosonic or fermionic) leads to fundamental collective quantum phenomena, namely Bose-Einstein condensation for bosons; and the formation of a Fermi surface for fermions \cite{PitaevskiiS2016,ashcroft1976}. From the point of view of quantum information, collective quantum behavior is generically associated with the appearance of entanglement among the elementary constituents \cite{Horodeckietal2009}. {Yet the notion of entanglement among identical quantum particles is complicated by their indistinguishability \cite{Benattietal2020}: even though the (anti-)symmetrization of the many-body wavefunction leads to formally entangled states, this entanglement between particles is only apparent \cite{Eckertetal2002,Ghirardietal2004}. For these systems a most natural form of entanglement is represented by \emph{mode} entanglement, i.e. the entanglement among the ``parties" (i.e. the subsets of modes) which share the particles \cite{WisemanV2003}. This form of entanglement is intrinsically connected with that of first-order coherence \cite{WisemanV2003,Macieszczaketal2019}, as well as higher-order coherence (when parties are not allowed to exchange particles). If the populations of single-particle modes are well defined (as in a Slater determinant for fermions, or in a Slater permanent for bosons), then entanglement is clearly absent among these modes. 
Nonetheless the picture may radically change when the mode basis is changed; and in particular when the modes are split (by passing them through beam splitters) \cite{Killoranetal2014, Morrisetal2020}. Entanglement among modes after splitting descends in fact from the indistinguishable nature of the particles occupying the modes, and it is therefore a characteristics (not a limitation) of identical particles. Hence entanglement between identical particles can be turned to mode entanglement; and moreover one can show that it represents a resource \cite{Morrisetal2020}, allowing a quantum state of identical particles to outperform separable states of distinguishable particles for specific tasks.

A relevant example is offered by particles with an internal structure, akin to a spin. For ensembles of indistinguishable spinful particles, it is meaningful to consider the entanglement among their internal degrees of freedom -- hereafter denoted \emph{spin entanglement} -- as if the spins were carried by distinguishable particles. Indeed, upon splitting the spin ensemble into different parts, spin entanglement translates into spatial entanglement and Einstein-Podolsky-Rosen nonlocality, as shown by recent remarkable experiments on ultracold quantum gases \cite{fadeletal2018,kunkeletal2018,langeetal2018,Colciaghietal2023}. Moreover spin entanglement realized in systems of indistinguishable particles (such as the one associated to spin-squeezing \cite{Ma2011PR,Pezze2018RMP}) is at the core of the increased sensitivity of spin ensembles to external fields with respect to the case of uncorrelated spins \cite{Pezze2018RMP, Morrisetal2020}. And this metrological advantage stemming from entanglement may be completely agnostic to whether the spins are distinguishable or not. In the following we shall adopt this theoretical framework to study entanglement of indistinguishable particles, and we consider their spins on the same footing as those of distinguishable particles.} 

In this work we explore the effect of quantum degeneracy of ideal $S=1/2$ quantum gases on their spin state, and specifically on its entanglement properties. Previous theoretical works on spinful ideal gases have studied the entanglement properties of the reduced state of two particles \cite{Vedral2003, OhKim2004}. Here we focus instead our attention on entanglement witnessing \cite{Guehne_2009} based on the fluctuation properties of the \emph{collective spin} of the spin ensemble. When the knowledge of the first and second moments of these fluctuations is accessible -- as in state-of-the-art experiments on ultracold spin ensembles \cite{Riedel2010,Luecke2011,Muessel2014PRL,schmiedetal2016,Hosten2016,fadeletal2018,kunkeletal2018,langeetal2018,Quetal2020} --  the criteria of Ref.~\cite{Toth2009PRA} (generalized in Ref.~\cite{Hyllusetal2012} to the case of fluctuating particle numbers) allow one to conclude whether this knowledge is sufficient or not to witness the presence of multipartite entanglement in the system. In our work we show that the knowledge of the variance of the collective spin fluctuations is insufficient to witness {spin} entanglement in ideal spinful Bose gases. For ideal Fermi gases, instead, the ground state of a spin-balanced mixture realizes a total spin singlet, whose multipartite entanglement is witnessed by the variance criteria. We establish the parameters (temperature, spin imbalance) in which entanglement due to singlet formation can be witnessed for an ideal Fermi gas in free space and in a parabolic trap. In both cases detectable spin entanglement appears when the gas enters the quantum-degenerate regime. We then inspect the real-space spin-spin correlations associated with singlet formation for a lattice gas of fermions, and show that, while these correlations have an extended spatial structure, they are not sufficient to witness entanglement among different sites of a lattice. Our findings highlight the intimate connection between Fermi statistics and spin entanglement, which offers a direct probe of the superposition nature of antisymmetrized states; and they can be tested experimentally by using state-of-the-art manipulation and detection techniques for the collective spin of ultracold atoms.  

\section{Collective-spin fluctuations in ideal quantum gases}

We consider a system of non-interacting quantum particles, with single-particle Hamiltonian possessing eigenvectors $|\psi_\alpha\rangle$ and related eigenstates $\epsilon_\alpha$. In second quantization   
\begin{equation}
{\cal H} = \sum_{\alpha \sigma} \left (  \epsilon_\alpha - \sigma H \right ) a^\dagger_{\alpha \sigma} a_{\alpha\sigma}
\end{equation}
where $a_{\alpha\sigma}$ ($a^\dagger_{\alpha\sigma}$) destroys (creates) a particle in the motional state $|\psi_\alpha\rangle$ and with spin state $|\sigma\rangle$. Here we shall assume that there is no spin-orbit coupling for single particles, such that the only dependence of the energy on the spin is through an applied magnetic field $H$ controlling the spin polarization.  In the following the operators $a_{\alpha,\sigma}$, $a^\dagger_{\alpha,\sigma}$ will be either bosonic of fermionic, namely $a_{\alpha,\sigma}a^\dagger_{\alpha',\sigma'} - \eta ~a^\dagger_{\alpha',\sigma'} a_{\alpha,\sigma}  = \delta_{\alpha,\alpha'} \delta_{\sigma,\sigma'}$ and $a_{\alpha,\sigma}a_{\alpha',\sigma'} - \eta ~a_{\alpha',\sigma'} a_{\alpha,\sigma} = 0$, with $\eta = 1$ for bosons and $\eta = -1$ for fermions. 

The equilibrium state of the ideal quantum gas in the grand-canonical ensemble, with temperature $T$ and chemical potential $\mu$, is a Gaussian state $\rho = e^{-\beta ({\cal H}-\mu N)}/Z_G$; with 
$\beta = 1/(k_B T)$, $N = \sum_{\alpha,\sigma} a^\dagger_{\alpha\sigma} a_{\alpha\sigma}$ is the number operator; and $Z_G = {\rm Tr}(e^{-\beta ({\cal H}-\mu N)})$ is the grand-canonical partition function. Such a state satisfies Wick's theorem, implying in particular for a product of four field operators 
\begin{align}
& \langle a_{k_1}^\dagger a_{k_2} a_{k_3}^\dagger a_{k_4} \rangle = \\
& \langle a_{k_1}^\dagger a_{k_2} \rangle \langle a_{k_3}^\dagger a_{k_4} \rangle + \langle a_{k_1}^\dagger a_{k_4} \rangle \left  ( \delta_{k_2,k_3} + \eta \langle a_{k_3}^\dagger a_{k_2} \rangle \right )
\nonumber
\end{align}
where $k_i = (\alpha_i \sigma_i)$ denotes a single-particle motional and spin state. 

The collective-spin operators of the spin ensemble takes then the following quadratic expression in terms of field operators 
\begin{equation}
J^\mu = \sum_\alpha \sum_{\sigma \sigma'} a^\dagger_{\alpha\sigma} \tau^\mu_{\sigma\sigma'} a_{\alpha\sigma'}
\label{e.Jmu}
 \end{equation}
 where $\mu = x,y,z$ and $\tau^\mu$ are spin matrices providing a representation of the SU(2) group. In the following we shall specialize to the case of $S=1/2$ spins, so that $\tau^\mu =  \sigma^\mu/2$ where $\sigma^\mu$ are Pauli matrices; but the treatment we offer here can be readily generalized to larger spins. 
 
 Making use of the above-cited Wick's theorem, the variances of the fluctuations of the collective spin components, ${\rm Var}(J^\mu) = \langle (J^\mu)^2\rangle - \langle J^\mu\rangle^2$, take then the following simple forms
 \begin{eqnarray}
 {\rm Var}(J^z) &=&  \frac{\langle N \rangle}{4} +  \frac{\eta }{4} \sum_\alpha \left ( n^2_{\alpha \uparrow} + n^2_{\alpha \downarrow} \right ) \label{e.Varz} \\
 {\rm Var}(J^{x(y)}) & = & \frac{\langle N \rangle}{4} +  \frac{\eta }{2} \sum_\alpha n_{\alpha \uparrow} n_{\alpha \downarrow}~ \label{e.Varxy}
 \end{eqnarray}
 where $n_{\alpha \sigma} =  \langle a^\dagger_{\alpha\sigma} a_{\alpha\sigma} \rangle = \{ \exp[\beta (\epsilon_{\alpha} - \sigma H - \mu)] - \eta \}^{-1}$ is the Bose-Einstein ($\eta = 1$) of Fermi-Dirac ($\eta=-1$) distribution. 
 In particular ${\rm Var}(J^z)  = \langle (J^z)^2 \rangle - \langle N_{\uparrow}- N_{\downarrow} \rangle^2/4$ with $N_\sigma = \sum_{\alpha} a^\dagger_{\alpha\sigma} a_{\alpha\sigma}$ the number operator for particles with spin $\sigma = \uparrow, \downarrow$; while  ${\rm Var}(J^{x(y)}) =  \langle (J^{x(y)})^2 \rangle$, as we choose (without loss of generality) that a net spin polarization only appears along the $z$ direction. 
 
\section{Entanglement criteria from the variance of spin fluctuations} 
\label{s.entanglement}

The knowledge of the variance of the collective spin fluctuations is a crucial piece of information when aiming at establishing the entangled nature of the quantum state of a spin ensemble. In particular it allows one to estimate the so-called entanglement depth \cite{SorensenM2001}, namely the amount of spins which are involved in the entangled state -- as shown by a significant number of recent experiments \cite{Riedel2010,Luecke2011,Lueckeetal2014,Zouetal2018}. The ability to witness entanglement via the variance of the collective spin fluctuations relies on the existence of an exhaustive set of criteria first established by Ref.~\cite{Toth2009PRA} in the case of a well defined number of $S=1/2$ spins. Testing the data on the variances of spin fluctuations against these criteria allows one to conclude whether the knowledge on the variances is sufficient or not to witness the presence of entanglement among the spins.  
 These criteria have been generalized by Ref.~\cite{Hyllusetal2012} to the case of a fluctuating number of particles, and this is the version we shall adopt in this work, since we consider ideal gases in the grand-canonical ensemble. Ref.~\cite{Hyllusetal2012} has established in particular that all separable states with a fluctuating particle number
 \begin{equation}
\rho_{\rm sep} = \sum_N P_N \sum_s p_s \otimes_{i=1}^N \rho_i^{(s)}~,
\label{e.sep}
 \end{equation}
 with particle-number distribution $P_N$ such that $P_0 = P_1 = 0$, satisfy the following inequalities
 \begin{align}
 & {\rm Var}(J^x) + {\rm Var}(J^y) + {\rm Var}(J^z) \geq  \frac{\langle N \rangle}{2} \label{e.1} \\
 & {\rm Var}(J^{\mu_1}) \geq \left \langle \frac{1}{N-1} \left [ (J^{\mu_2})^2 +J^{\mu_3})^2 \right ] \right \rangle - \left \langle \frac{N}{2(N-1)} \right \rangle \label{e.2} \\
 & {\rm Var}(J^{\mu_1}) +  {\rm Var}(J^{\mu_2}) \geq  \left \langle \frac{1}{N-1} (J^{\mu_3})^2  \right \rangle + \left \langle \frac{N(N-2)}{4(N-1)} \right \rangle \label{e.3}
 \end{align}
 where $\mu_1$, $\mu_2$ and $\mu_3$ are any permutation of $x$,$y$ and $z$. 
 Therefore the violation of any of these three inequalities contradicts the separable form Eq.~\eqref{e.sep} for the state, and it witnesses the presence of so-called \emph{particle entanglement}. 
 
 When specializing one's attention to the case of quantum gases consisting of indistinguishable particles, the separable form Eq.~\eqref{e.sep} is compatible with a bosonic gas only if $\rho_i^{(s)} = \rho^{(s)}$ for all particles, such that the state is a statistical mixtures of states which are fully symmetric under particle exchange. In the case of fermions, Eq.~\eqref{e.sep} is incompatible with the anti-symmetrization principle (unless one has only a single particle). Therefore Fermi-Dirac statistics comes inevitably with particle entanglement. 
 
  In light of the above observation, one may question the relevance of witnessing entanglement in the case of fermions. Yet one can argue that phenomena which are exclusively related to the antisymmetrization principle, and not obtainable otherwise, are not necessarily obvious. As an example, the very Fermi-Dirac statistics of single-state occupations is certainly a consequence of the antisymmetric nature of fermionic many-particle states; but, once a specific single-particle basis is chosen, this very statistics could be in principle mimicked using a separable state as in Eq.~\eqref{e.sep}. Hence observing a Fermi-Dirac distribution is not per se evidence of entanglement. Having instead observations that are incompatible with \emph{any} separable state -- as \emph{e.g.} by violating any of the inequalities Eqs.~\eqref{e.1}, \eqref{e.2} and \eqref{e.3} --  is a direct observation of the intimately entangled nature of fermionic states, and their radical departure from any possible picture contaning only classical correlations among particles.

\section{Ideal Bose gas and absence of entanglement witnessing}

We start our discussion on spin-entanglement witnessing via Eqs.~\eqref{e.1}, \eqref{e.2} and \eqref{e.3} from the case of the ideal Bose gas. For that case we have that
${\rm Var}(J^\mu) > \langle N \rangle/4$ for all $\mu = x,y,z$, following from Eqs.~\eqref{e.Varz} and \eqref{e.Varxy} with $\eta = 1$. 
As a consequence Eq.~\eqref{e.1} is always satisfied, since  $\sum_{\mu} {\rm Var}(J^\mu) \geq 3\langle N \rangle/4$. 

As for the inequality Eq.~\eqref{e.3}, one has that $(J^{\mu_3})^2 \leq N^2/4$ for any fixed particle number $N$. As a consequence the right-hand side obeys the inequality
\begin{align}
& \left \langle \frac{1}{N-1} (J^{\mu_3})^2 \right \rangle + \left \langle \frac{N(N-2)}{4(N-1)} \right \rangle \nonumber \\ 
& \leq  \left \langle \frac{N^2}{4(N-1)} + \frac{N(N-2)}{4(N-1)} \right \rangle = \frac{\langle N \rangle}{2}
\end{align}
while the left-hand side obeys the inequality ${\rm Var}(J^{\mu_1}) + {\rm Var}(J^{\mu_2}) \geq \langle N \rangle/2$ (from Eqs.~\eqref{e.Varz} and \eqref{e.Varxy}). Therefore the inequality Eq.~\eqref{e.3} is always satisfied.

As for the inequality Eq.~\eqref{e.2}, it should be examined (combined with Eq.~\eqref{e.3}) in two different cases:
\begin{itemize}
\item $\mu_1 = z, \mu_2 = x, \mu_3 = y$. In this case one can subtract Eq.~\eqref{e.2} from twice Eq.~\eqref{e.3} to get 
\begin{equation}
{\rm Var}(J^z) + 2 ~{\rm Var}(J^x) \geq \left \langle \frac{N(N-2)}{2(N-1)} + \frac{N}{2(N-1)} \right \rangle =  \frac{\langle N \rangle}{2}
\end{equation}
which is automatically satisfied, since in fact ${\rm Var}(J^z) + 2 ~{\rm Var}(J^x) \geq 3\langle N\rangle/4$. 

\item $\mu_1 = x(y), \mu_2 = y(x), \mu_3 = z$. In this case, subtracting again Eq.~\eqref{e.2} from Eq.~\eqref{e.3} we obtain
\begin{eqnarray}
 \left \langle \left ( 1+\frac{1}{N-1} \right ) (J^x)^2 \right \rangle & \geq &   \left \langle \frac{N(N-2)}{4(N-1)} + \frac{N}{2(N-1)} \right \rangle \nonumber \\  
 & = &\left \langle  \frac{N^2}{4(N-1)} \right \rangle \approx \left \langle  \frac{N}{4} \right \rangle
\end{eqnarray} 
which, by neglecting again $1/(N-1)$ with respect to 1, is satisfied by Eq.~\eqref{e.Varxy}. 
\end{itemize}

Therefore we can conclude that $S=1/2$ ideal Bose gases do not admit entanglement witnessing from the knowledge of the variances of collective spin fluctuations. 
Nonetheless one should not deduce from this result that the state of a spinful ideal Bose gas is always compatible with a separable state, but only that spin entanglement in ideal Bose gases can only be detected by using more sophisticated criteria. 

\section{Singlet formation in ideal Fermi gases}

In the case of an ideal Fermi gas, the situation is radically different.  Indeed, since $\eta = -1$, Eqs.\eqref{e.Varz} and \eqref{e.Varxy} imply that ${\rm Var}(J^\mu) \leq \langle N \rangle/4$, so that the inequality Eq.~\eqref{e.1} verified by all separable states can be violated, and spin entanglement can be detected using the variance of the collective spin fluctuations. 

In the absence of spin imbalance (namely for $H=0$), we have that $n_{\alpha\uparrow} = n_{\alpha\downarrow}$, so that ${\rm Var}(J^z) =  {\rm Var}(J^{x(y)})$ from Eqs.~\eqref{e.Varz} and \eqref{e.Varxy}. In the ground state at $T=0$, we have that $n_{\alpha\sigma} = 1$ for $\epsilon_\alpha \leq \mu$, and $n_{\alpha\sigma} = 0$ otherwise. Therefore $n^2_{\alpha\sigma} = n_{\alpha\sigma} n_{\alpha,-\sigma} = n_{\alpha\sigma}$, and $\sum_\alpha \left ( n^2_{\alpha\uparrow} + n^2_{\alpha\downarrow} \right) = 2 \sum n_{\alpha\uparrow} n_{\alpha\downarrow} = \langle N \rangle$, so that ${\rm Var}(J^z) = {\rm Var}(J^{x(y)}) = 0$. Since $\langle J^\mu\rangle = 0$ (in the absence of polarizing fields), this implies that $\langle {\bm J}^2 \rangle  = 0$, namely the ground state of the spinful ideal Fermi gas is a total spin singlet. This state offers the maximal violation of the inequality Eq.~\eqref{e.1}. 

The total singlet realized in the ground state can be simply thought of as a collection of two-fermion singlets formed by pairs of $\uparrow$ and $\downarrow$ particles occupying the same single-particle eigenstate $|\psi_\alpha\rangle$. If the single-particle eigenstates are spatially delocalized, this implies that spin entanglement leads to pronounced non-local correlations in the system. The two-fermion singlet formation may erroneously suggest that fermions are only entangled in pairs; indeed, if the pairs are spatially overlapping, one cannot factor the state of the system into pairs, $\rho \neq \sum_s p_s \otimes_{\rm pair} \rho_{\rm pair}^{(2)}$, as the latter state does not posses the right anti-symmetry upon exchange.  This means that the total singlet realized in the ground state has an entanglement depth (= number of particles forming a joint entangled state) which is generally larger than two; and which may be roughly estimated as the number of particles contained in the overlap region between two single-particle eigenstates $|\psi_\alpha\rangle$. Only if the eigenstates are completely separated spatially (such as for atoms in a very deep optical lattice) the entanglement depth would reduce to two particles.  

When introducing a finite temperature, or a finite polarizing field $H$, some of the singlet pairs will be broken, leading to a finite collective-spin length $\langle {\bm J}^2 \rangle$.  To evaluate how many particles are forming singlets, we can introduce a \emph{singlet-formation parameter}, introduced in Refs.~\cite{TothM2010,Behboodetal2014} 
\begin{equation}
\xi_s^2 = \frac{2 \sum_\mu {\rm Var}(J^\mu)}{\langle N \rangle}
\end{equation}
such that $\xi_s^2<1$ is a witness of entanglement via singlet formation -- as it implies the violation of Eq.~\eqref{e.1}. In particular 
\begin{equation}
f_s = 1-\xi_s^2
\end{equation}
 gives the effective fraction of particles which are involved in a spin singlet \cite{TothM2010,Behboodetal2014}. 

It is already intuitive that, as long as the state of the system retains Fermi degeneracy at finite $T$ (i.e. the Fermi-Dirac distribution is not completely smeared out), and it is not fully polarized by an external field, it should also retain spin entanglement from the formation of singlets resulting from the double occupancy of motional states $|\psi_\alpha\rangle$. In the following we shall make this intuition quantitative, by evaluating the effective fraction $f_s$ of fermions forming singlets across the phase diagram of the system in the case of a Fermi gas in free space, as well as in a parabolic trap.

\begin{figure*}[ht!]
\begin{center}
\includegraphics[width=0.9\textwidth]{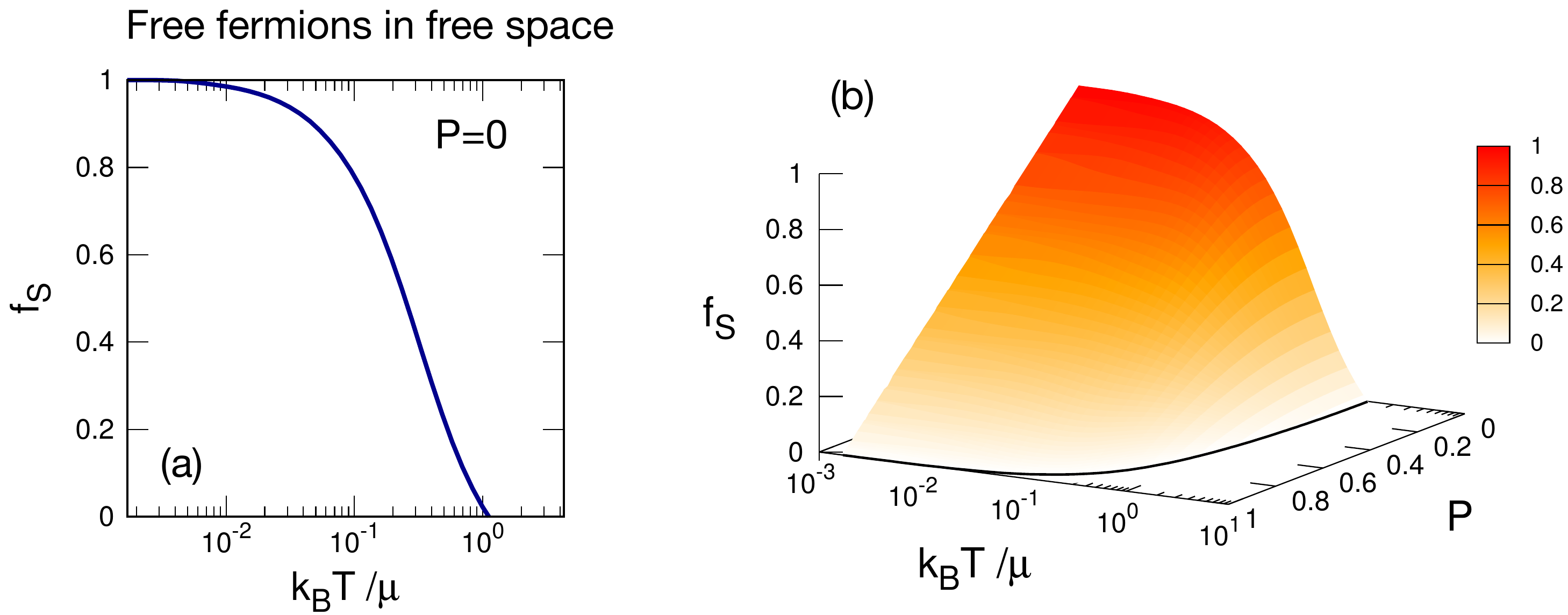}
\caption{Fraction of spin singlets for an ideal $S=1/2$ Fermi gas in 3d free space. (a) $f_s$ vs. temperature (in units of the chemical potential) at zero polarization $P$; (b) evolution of $f_s$ upon increasing polarization from 0 to 1. The calculation has been performed on a $30^3$ grid, which provides a good approximation to the thermodynamic limit.}
\label{f.freespace}
\end{center}
\end{figure*}

\subsection{Free space}
In the case of fermions in a 3d box of side $L$ with periodic boundary conditions,  $\epsilon_{\alpha} = \frac{\hbar^2 {\bm k}^2}{2m}$ with ${\bm  k} = \frac{2\pi}{L} (n_x,n_y,n_z)$ and $n_\mu \in \mathbb{Z}$. For $L\to \infty$ the populations $n_{\alpha\sigma}$ are just functions of $k_B T/\mu$ and of the applied magnetic field $H$ or, conversely, of the polarization $P = \langle N_{\uparrow} - N_{\downarrow} \rangle/ \langle N \rangle$; and therefore so are the variances of the collective-spin fluctuations and the parameter $f_s$. This parameter is shown in Fig.~\ref{f.freespace}(a) at $P=0$ as a function of $k_B T/\mu$, and it is found to vanish at $k_B T/\mu \approx 1.12$. Hence we clearly see that quantum degeneracy (namely the regime $k_B T \lesssim \mu$) implies the onset of spin entanglement as detectable via the variances of collective-spin fluctuations. 

Upon polarizing the Fermi gas the fraction of singlets is progressively reduced at all temperatures, and the range of $k_B T/\mu$ over which singlet formation is detected via the variances shrinks. Yet the decrease of the singlet fraction $f_s$ is linear in $P$ (as simply expected via pair breaking) only at low $T$, while it becomes non-linear -- and therefore much slower -- at intermediate temperatures of order $\mu$. This shows that spin entanglement revealed by the $f_s$ parameter can be robust to fluctuations in the polarization around $P=0$, which are inevitable in real experiments.  

\begin{figure*}[ht!]
\begin{center}
\includegraphics[width=0.9\textwidth]{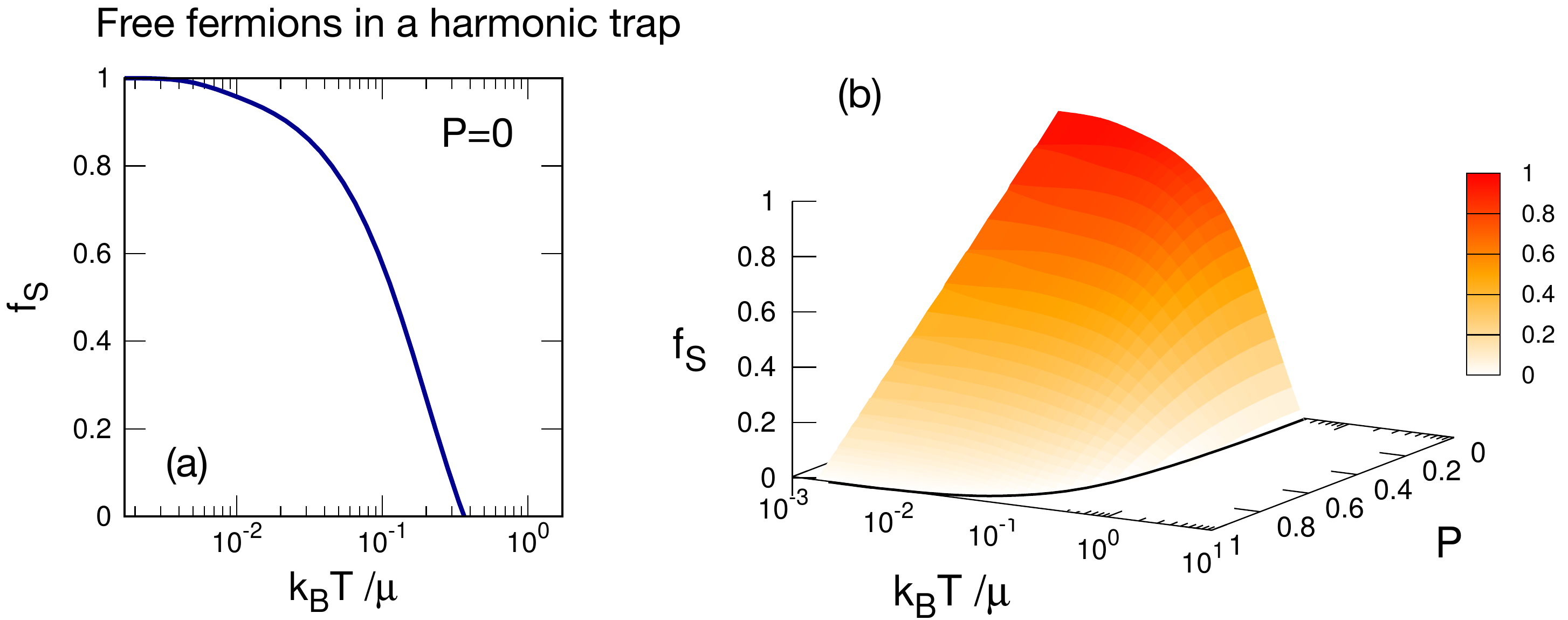}
\caption{Fraction of spin singlets for an ideal $S=1/2$ Fermi gas in a 3d harmonic trap with frequency $\hbar \omega = \mu/30$~. (a) $f_s$ vs. temperature (in units of the chemical potential) at zero polarization $P$; (b) evolution of $f_s$ upon increasing polarization from 0 to 1.}
\label{f.trap}
\end{center}
\end{figure*}

\subsection{Parabolic trap}

In order to come closer to the reality of ultracold-atom experiments, we have repeated the same calculation for the case of the Fermi gas in a harmonic trap with eigenenergies $\epsilon_\alpha = \hbar\omega( n_x + n_y + n_z + 3/2)$  ($n_\mu \in \mathbb{N}$). We have chosen a frequency $\mu / \hbar\omega = 30$, such that $\langle N \rangle \approx 1-2 \times 10^4$ at low temperature. The results of Fig.~\ref{f.trap} show again that the variances of collective-spin fluctuations witness the formation of spin singlets when degeneracy sets in, namely when $k_B T /\mu \lesssim 1$ (specifically $k_B T /\mu \leq 0.368$ at $P=0$ in this example); and that this witnessing is robust to polarization fluctuations close to the temperature at which $f_s$ vanishes.

\begin{figure*}[ht!]
\begin{center}
\includegraphics[width=0.9\textwidth]{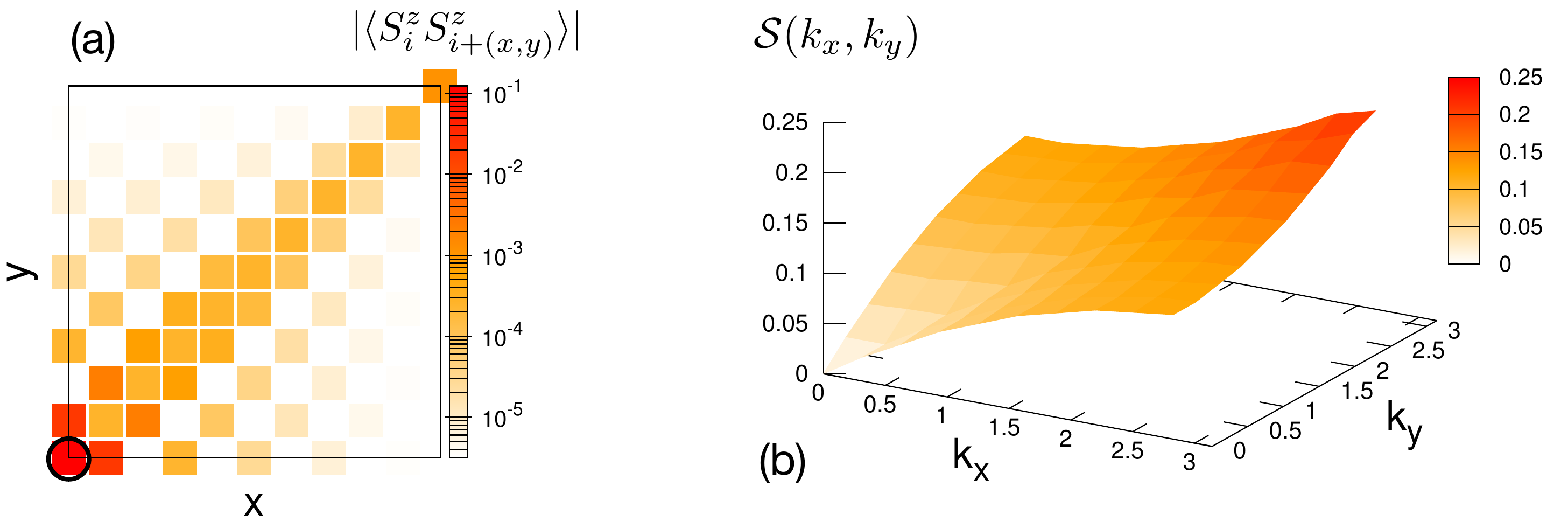}
\caption{Spin-spin correlations in the free Fermi gas on a square lattice. (a) Spin-spin correlations between the reference site (marked by the black circle) and the rest of the lattice; (b) spin structure factor.}
\label{f.correlations}
\end{center}
\end{figure*}

\section{Structure of spatial correlations on a lattice and entanglement witnessing}  

In this final section we address the consequences that singlet formation has on the spatial structure of spin-spin correlations, in the case in which singlet pairs are delocalized across the lattice; and the question whether the knowledge of spin-spin correlations leads to alternative entanglement-witnessing schemes. 

To examine this aspect, we discretize space, and consider a spinful Fermi gas on a 2d square lattice, with Hamiltonian
\begin{equation}
{\cal H} = -J \sum_{\langle ij \rangle, \sigma} \left (  a_{i\sigma}^\dagger a_{j\sigma} + {\rm h.c.} \right ) ~ 
\end{equation} 
where $\langle ij \rangle$ represent nearest-neighbor pairs of sites on a the square lattice. 
We consider a system at half filling, namely with vanishing chemical potential $\mu=0$. In this case, the ground state has fermions occupy in pairs the Bloch waves of the lattice, filling half of the Brillouin zone, and forming therefore spatially extended singlets. These singlets are in turn expected to induce negative spin-spin correlations at long distances. 

We calculate the spatial spin-spin correlation function for this system, which reads 
\begin{equation}
\langle S^z_i S^z_j \rangle = \frac{1}{4} \left \langle \left ( a_{i\uparrow}^\dagger a_{i\uparrow} - a_{i\downarrow}^\dagger a_{i\downarrow}  \right) \left (a_{j\uparrow}^\dagger a_{j\uparrow} - a_{j\downarrow}^\dagger a_{j\downarrow}  \right) \right \rangle
\end{equation}
where we have introduced the on-site spin operators $S_i^\mu = \sum_{\sigma\sigma'} \tau^\mu_{\sigma \sigma'} a_{i\sigma}^\dagger a_{i\sigma'}$, in analogy with the definition of  Eq.~\eqref{e.Jmu}. Similar to ordinary $S=1/2$ operators, these operators have a spectrum bounded between $-1/2$ and $1/2$; yet, unlike ordinary spin operators, they admit also the eigenvalue 0, corresponding to the case of an empty site, or of a doubly occupied site (with two fermions forming a local singlet).

When evaluated on the equilibrium state of a free Fermi gas, the spin-spin correlation function takes the form (via Wick's theorem)
 \begin{equation}
\langle S^z_i S^z_j \rangle = \begin{cases} \frac{1}{4} \sum_\sigma n_{i\sigma} (1-n_{i\sigma}) ~~~~ i =j\\
- \frac{1}{4} \sum_{\sigma} |G^{(\sigma)}_{ij}|^2  ~~~~ i \neq j
\end{cases}
\end{equation}
where we introduced the {first-order correlation} function $G^{(\sigma)}_{ij} = \langle a^\dagger_{i\sigma} a_{j\sigma}\rangle$, which, for a gas on a translationally invariant $L\times L$ lattice, is simply the (inverse) Fourier transform of the momentum-space distribution $n_{\bm k\sigma} = \langle a_{\bm k\sigma}^\dagger a_{\bm k\sigma}\rangle$ 
\begin{equation}
G^{(\sigma)}_{ij} = \frac{1}{L^2} \sum_{\bm k} e^{-i \bm k \cdot (\bm r_i - \bm r_j)} ~n_{\bm k\sigma}~.
\end{equation}
Hence we observe that offsite spin-spin correlations are negative at all distances -- as expected from the presence of delocalized singlets -- and they are spatially modulated by the inverse Fourier transform of the momentum-space distribution, namely of a step function with support on half the Brillouin zone in the ground state.  
A picture of the correlation pattern is provided in Fig.~\ref{f.correlations}(a): there we observe that correlations have a significant spatial extent as resulting from the delocalized nature of singlets; and a characteristic spatial modulation at short wavelength which is inherited from the oscillating behavior of the {first-order correlation} function, occurring at twice the wavevectors at the Fermi surface. Yet the correlations are rather tenuous, and become extremely weak beyond nearest neighbors. 

As a result, the Fourier transform of the spin-spin correlations, namely the spin structure factor
\begin{equation}
{\cal S}(k_x, k_y) = \frac{1}{L^2} \sum_{ij} e^{i{\bm k}(\bm r_i - \bm r_j)} \langle S_i^z S_j^z \rangle
\end{equation}
does not exhibit any sharp peak, but only a weak maximum at ${\bm k} = (\pi,\pi)$. This maximum stays below the value of 1/4, which means that it does not rise above the maximum value that the structure factor takes for an ensemble of uncorrelated, localized $S=1/2$ spins on the square lattice. This means that spin-spin correlations are not sufficiently strong to enable witnessing of \emph{spatial} spin entanglement (among the spins located at different sites) from the inspection of the structure factor. One could formalize this statement by introducing the quantum Fisher information (QFI) \cite{Pezze2018RMP} related to the staggered magnetization $J^z_{\rm stag} = \sum_i (-1)^i S_i^z$, {which in the ground state coincides with four times the variance,  ${\rm QFI}(J^z_{\rm stag}) = 4  {\rm Var}(J^z_{\rm stag}) = 4 L^2 {\cal S}(\pi,\pi)$. The QFI of a collective spin operator such as $J^z_{\rm stag}$ serves as a witness of multipartite entanglement whenever its density exceeds unity \cite{Toth2012, Hyllus2012PRA}. The fact that ${\rm QFI}(J^z_{\rm stag})/L^2 = 4 {\cal S}(\pi,\pi) < 1$ in the ground state of the ideal Fermi gas implies that the quantum fluctuations of the staggered collective spin are not sufficiently large to witness spatial multipartite entanglement in this system of distributed $S=1/2$ spins.} The most salient feature of the structure factor remains instead the fact that ${\cal S}(0,0) = 0$, reflecting the fact that the system forms a perfect singlet in the ground state -- and this aspect allows one to witness particle entanglement (not spatial entanglement), as already commented in Sec.~\ref{s.entanglement}.

\section{Conclusions}

In this work we have explored the consequences of quantum degeneracy on the appearance of spin entanglement in ideal spinful quantum gases. We showed that ideal Bose gases do not possess spin entanglement detectable by the knowledge of the variances of collective-spin fluctuations; while, due to the formation of spin singlets, ideal Fermi gases develop spin entanglement which is revealed by a strong suppression of the variance of all collective spin components. Spin entanglement in Fermi gases is a consequence of the anti-symmetrization principle, and a direct probe of the superposition nature of the quantum states of indistinguishable particles. 

While singlet formation leads to the appearance of spatial spin-spin correlations, we have shown that multipartite entanglement among spatial modes in a Fermi lattice gas is not witnessed by such correlations when using the criterion based on the quantum Fisher information. In order to detect the link between spin correlations and entanglement via the quantum Fisher information it is necessary to move beyond the ideal Fermi gas picture, and to introduce interactions. In the limit of very strong interactions on a bipartite lattice, a Fermi gas at half filling realizes Heisenberg antiferromagnetism \cite{Mazurenko2017}: this is associated with the maximal delocalization of spin singlets resulting in long-range spin-spin correlations, which clearly exhibit multipartite entanglement.  It will be therefore interesting to study in the future the interaction strength required to stabilize spin-spin correlations which can be unambiguously associated with multipartite entanglement.   

The formation of total spin singlets in a spin ensemble is a very remarkable fact, implying that quantum noise of the collective spin is completely silenced. Total spin singlets are not only highly entangled, but they also have fundamental properties of Bell nonlocality \cite{FrerotR2021,Mueller-Rigatetal2021}. The reduction of spin noise due to singlet formation has been realized experimentally using Bose gases, via protocols of measurement-induced spin entanglement \cite{TothM2010, Behboodetal2014,Kong2020}, as well as of adiabatic preparation in the presence of antiferromagnetic all-to-all interactions \cite{Evrardetal2021}.  In the case of Fermi gases, the suppression of spin noise due to pairing and/or strong interactions has been studied theoretically \cite{Bruunetal2009}, and observed experimentally, in relationship with Fermi degeneracy and strong interactions in Ref.~\cite{Meinekeetal2012}, leading to the witnessing of spin entanglement along the lines that we discussed in this work. In comparison with the previous studies, our work establishes a link between the spin-noise suppression and the formation of entangled singlets; and it clarifies the stability of singlet formation in the case of non-interacting Fermi gases in relationship with quantum degeneracy and spin imbalance.    
Interesting extensions of our work would imply the study of degenerate quantum gases carrying a spin larger than $S=1/2$, as well as gases of SU(N) fermions \cite{Cazalilla2014}. 

 Our findings are relevant to ongoing experiments on spin ensembles, and they would greatly profit from the ever-increasing sensitivity of such experiments to the collective-spin fluctuations, down to the level of fundamental quantum noise \cite{Quetal2020}. Indeed singlet formation in spin-balanced Fermi gases implies the complete suppression of all collective-spin fluctuations, so that its faithful detection in an experiment requires in principle the ultimate level of sensitivity. From the point of view of metrology, singlet states are insensitive to uniform magnetic fields, but remain sensitive to magnetic-field gradients, as discussed in Ref.~\cite{Urizar-Lanzetal2013}. Yet it is not obvious whether the entangled nature of total spin singlets leads to a metrological advantage compared to separable states, and the spatial structure of the spin singlet plays an important role in this respect. As an example, our results on quantum Fisher information of the ideal Fermi gas on a lattice show that there is no metrological advantage in field sensing by using a spin-entangled ideal Fermi gas compared to a system of localized and uncorrelated spins. As already mentioned above, interactions in the Fermi gas would be needed instead in order to stabilize correlations leading to entanglement-enhanced metrological sensitivity.

\begin{acknowledgements}
Fruitful discussions with M. Robert-de-Saint-Vincent are gratefully acknowledged. 
The work of TR is supported ANR (EELS project), QuantERA (MAQS project), and PEPR-q (QubitAF project).   FZA acknowledges the support of the Algerian Ministry of Higher Education and Scientific Research, and the ENS of Lyon for hospitality. 
\end{acknowledgements}

\newpage

\bibliography{singlets.bib}

\end{document}